\documentclass[%
 reprint,
 amsmath,amssymb,
 aps,
]{revtex4-2}

\usepackage{soul}
\usepackage{xcolor}
\usepackage{graphicx}
\usepackage{dcolumn}
\usepackage{bm}
\usepackage{tikz}

\begin{document}

\preprint{APS/123-QED}

\title{Neutrino monitoring of explosions for excluding fission yield}

\author{O. Benevides Rodrigues$^{2}$}
\author{N. S. Bowden$^{4}$}
\author{R. Carr$^{8}$}
\author{A. Conant$^{6}$}
\author{M. Foxe$^{7}$}
\author{D. Hornback$^{1}$}
\author{P. Huber$^{10}$}
\author{A.~Irani$^{2}$}
\author{L. Lebanowski$^{3,9}$}
\author{V. A. Li$^{4}$}
\author{J. M. Link$^{10}$}
\author{B. R. Littlejohn$^{2}$}
\author{F. Machado$^{2}$}
\author{M. P. Mendenhall$^{4}$}
\author{H.~P. Mumm$^{5}$}
\author{J. Newby$^{6}$}
\author{I. D. Olusola$^{10}$}
\author{G. D. Orebi Gann$^{3,9}$}
\author{T. Papatyi$^{7}$}
\author{L. Pickard$^{3,9}$}
\author{X. Zhang$^{4}$}

\affiliation{$^{1}$Brookhaven National Laboratory, Upton, NY, USA}
\affiliation{$^{2}$Department of Physics, Illinois Institute of Technology, Chicago, IL, USA}
\affiliation{$^{3}$Lawrence Berkeley National Laboratory, Berkeley, USA}
\affiliation{$^{4}$Lawrence Livermore National Laboratory, Livermore, CA, USA}
\affiliation{$^{5}$National Institute of Standards and Technology, Gaithersburg, MD, USA}
\affiliation{$^{6}$Oak Ridge National Laboratory, Oak Ridge, TN, USA}
\affiliation{$^{7}$Pacific Northwest National Laboratory, Richland, WA, USA}
\affiliation{$^{8}$United States Naval Academy, Annapolis, MD, USA}
\affiliation{$^{9}$Department of Physics, University of California, Berkeley, CA, USA}
\affiliation{$^{10}$Center for Neutrino Physics, Virginia Tech, Blacksburg, VA, USA}

\date{\today}

\begin{abstract}

Nuclear fission produces neutrinos, so the absence of a neutrino signal can be used to set a limit on the fission content of an explosion. This capability could be employed on former nuclear test sites to assure regulators, international monitors, or other observers that activities involving chemical explosions do not exceed a designated limit for nuclear fission. This paper quantifies the neutrino detector masses that would be required to set fission yield limits at source-to-detector distances up to 100 km, assuming detection by inverse beta decay with realistic background levels. The analysis indicates that detectors with active mass in the ton- to tens-of-kiloton range can set potentially useful limits on the fission yield of large chemical explosions at the Nevada National Security Site. In contrast, inverse beta decay detectors are not well suited to excluding fission yield at longer range or in the subcritical nuclear experiments that have occurred at some test sites following the cessation of explosive nuclear testing. 

\end{abstract}

\maketitle


\section{\label{sec:Intro} Introduction}

\begin{figure*}[ht!]
    \centering
    \includegraphics[width=0.46\linewidth,trim=1.5cm 0.5cm 1cm 1cm]{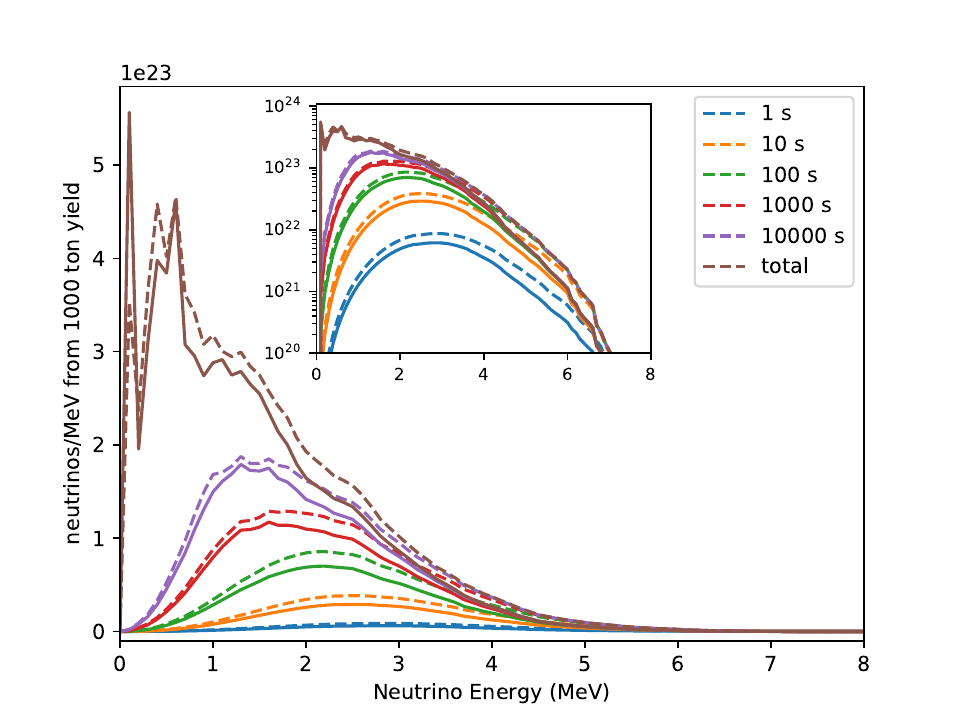}
    \hspace{0.05\linewidth}\includegraphics[width=0.46\linewidth,trim=1.5cm 0.5cm 1cm 1cm]{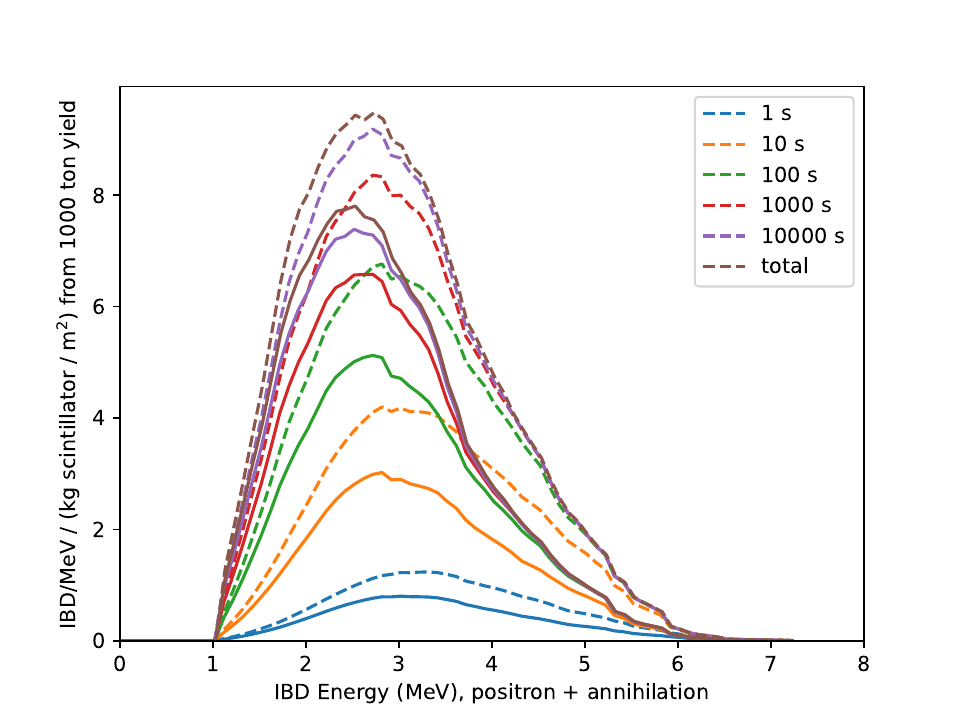}
    \caption{Left: Simulated energy spectrum of neutrinos released by an explosion of $^{235}$U (dashed) and $^{239}$Pu (solid), within successively longer time intervals, plotted as a function of neutrino energy. The inset plot shows the same data with a logarithmic vertical axis. Right: Fluxes from the left plot convolved with the inverse beta decay (IBD) cross section, plotted as a function of energy visible in an ideal IBD detector (the kinetic energy of the positron plus the energy of gammas released in the eventual positron + electron annihilation).}
    \label{fig:flux}
\end{figure*}

A nuclear fission explosion releases an intense burst of neutrinos, a distinctive signal that travels unimpeded through matter. Since the early years of nuclear weapons testing, physicists have considered possible uses for this signal in both basic science experiments and nuclear nonproliferation applications \cite{reines, LANL1980s, Bernstein:2019hix, annrev}. Most nonproliferation studies have focused on applications in which the neutrino detector would be hundreds or thousands of kilometers from potential fission sources. One study considered using neutrino detectors to search the globe for nuclear explosions below the threshold of the seismic, hydroacoustic, and infrasound sensors of the Comprehensive Nuclear-Test-Ban Treaty Organization's International Monitoring System \cite{Bernstein2001}. Another study considered using neutrinos as a complement to radioisotope detection in confirming the nuclear nature of weapons tests picked up via seismic sensors \cite{Carr2017}. These global-scale concepts exploit the fact that neutrinos easily penetrate the earth, but they face practical limitations due to the inverse square scaling of the neutrino flux with distance from a fission source.

This paper explores a concept which has received less attention in prior literature: using a neutrino detector to set limits on the fission yield of events occurring relatively close ($\lesssim 100$ km) to the detector. The idea is to collect neutrino data over a fixed time interval following an event and to use the absence of a detected signal within that window to place an upper limit on the amount of fission that could have occurred. In principle, the event could be anything that occurs at a known time and place and is brief in duration compared to the signal collection window. Since there is often special interest in demonstrating the non-fission nature of activities on former nuclear test sites, this paper concentrates on two types of events that have occurred on these sites since the early 1990s. The first type is chemical explosions with fairly large explosive yields, often designed to improve and validate capabilities for nuclear explosion monitoring. In the United States, the National Nuclear Security Administration (NNSA) has used chemical explosions with TNT-equivalent yields up to about 10 metric tons at the Nevada National Security Site (NNSS) \cite{SPE:2022, PE1:2024}. The second event type considered in this paper is subcritical experiments. These experiments use smaller chemical explosions to briefly subject plutonium or other special nuclear material to extreme heat and pressure characteristic of nuclear explosions. The United States conducts such experiments at the Principal Underground Laboratory for Subcritical Experimentation (PULSE), also at the NNSS.

Neither chemical explosions nor subcritical experiments produce a self-sustaining fission chain reaction, but their occurrence on former nuclear test sites has at times raised questions \cite{cnn2023, reuters2023}. In recent years, U.S. officials have taken steps to reassure external observers about the nature of test site activities \cite{nnsa_press_release_2024}. The neutrino concept developed in this paper could be relevant in that context or in other cases of confidence building, treaty verification, or regulatory oversight. Unlike the more conventional fission signatures of neutrons or gamma rays, neutrinos can be observed in a detector that is completely physically isolated from the monitored event. That arrangement may allow sensitive details to be kept more confidential, as is often desired in nuclear verification exercises. Only one publicly available study has discussed a use case along these lines, and it is limited to a basic scaling exercise for a single baseline and signal strength \cite{Bernstein2001}. This paper provides a more comprehensive analysis covering a range of sources and baselines, along with an empirically grounded treatment of backgrounds. 

The analysis begins in Section \ref{sec:source} with a model of the neutrino signal produced by a fission explosion. Section \ref{sec:detector} describes how that neutrino signal would be pursued in this application, keeping the assumptions about detector technology as generic as practical. Section \ref{sec:sensitivity} outlines a sensitivity framework that can be used to set limits on fission yield. Section \ref{sec:exclusion} discusses the minimum realizable standoff of a neutrino monitoring system from a chemical explosion given practical physical constraints. Section \ref{sec:results} presents sensitivity results, and Section \ref{sec:conclusions} concludes with some observations about the potential utility of neutrinos for test site monitoring scenarios.

\section{Signal Model}
\label{sec:source}

To assess how neutrino detectors can set limits on fission yield, it is necessary to understand the neutrino emission from a fission explosion. Like a nuclear reactor, a fission explosion produces neutrinos through the beta decay of nuclear fission fragments. These neutrinos are produced in the electron flavor and are, to be precise, antineutrinos. Following common practice, this paper uses neutrino as shorthand. The energies of neutrinos emitted from a fission explosion, as from a reactor, range up to about 10 MeV. Unlike in a reactor, all fissions in a typical nuclear explosion source occur within a microsecond \cite{Serber1992}. The neutrino emission from an explosion is therefore a sharp burst that trails off over a timescale set by the beta decay lifetimes of generated fission fragments. The average neutrino energy decreases with time, due to the anticorrelation between decay energy (Q value) and lifetime in beta decays.

Since different fissile materials produce different fission fragment distributions, the time and energy profiles of neutrinos emitted from a fission explosion depend on the isotope(s) fueling it. Historically, nuclear weapons and nuclear explosive tests have been fueled by $^{235}$U, $^{239}$Pu, or a mixture of isotopes that may also include $^{238}$U. Neutrinos from a fission explosion have never been observed, so producing a source term is a task for simulation. This analysis uses the CONFLUX package for that task \cite{Zhang:2025ayc}. Figure \ref{fig:flux} (left side) shows the simulated energy spectrum of neutrinos released by an explosion of $^{235}$U or $^{239}$Pu over time intervals ranging from 10 s to 10,000 s. In total, a kiloton of fission yield corresponds to the release of roughly $10^{24}$ neutrinos. (Here, and in the rest of this paper, ton refers to a metric ton, and tons of yield refers to tons of TNT-equivalent yield.) As Fig. \ref{fig:flux} shows, $^{239}$Pu produces fewer neutrinos per fission than $^{235}$U. (For simplicity, the plot does not show $^{238}$U, which produces more neutrinos per fission than both plotted isotopes.) In a monitoring scenario where the identity of any potential fission fuel is unknown, the most conservative fission yield limit will therefore be set by assuming that the explosion is fueled entirely by $^{239}$Pu. 
In this study, the neutrino source is assumed to be point-like, a reasonable approximation for the standoff distances and overall level of precision in this analysis. Although neutron activation of materials surrounding a fission explosion may contribute to the neutrino flux, that subdominant component is neglected in this study.

Since this paper aims to present a generalized analysis applicable to a range of neutrino detector technologies, assumptions about how the signal is detected are kept as minimal as possible. The primary assumption is that the detection channel is inverse beta decay (IBD), the interaction most commonly used to detect neutrinos from nuclear reactors. In this channel, a small fraction of neutrinos incident on a detector react with a free proton in hydrogen-containing target media to create a positron and unbound neutron:
\begin{equation}
    \bar{\nu}_e + p \rightarrow e^+ + n.
\end{equation}
This analysis uses a standard reference for the IBD cross section \cite{Vogel:1999zy}. Figure \ref{fig:flux} (right side) shows the flux from $^{235}$U and $^{239}$Pu explosions convolved with the IBD cross section, as it would be visible in an ideal IBD detector. The IBD interaction has a neutrino energy threshold of 1.8 MeV, and the energy visible in an ideal IBD detector is 0.8 MeV lower than the incident neutrino energy, which is why the cross-section-convolved flux begins at 1~MeV. While more than half of the total neutrino emission occurs below the IBD threshold, the sub-threshold fraction is lower at the early times targeted in this analysis.

\begin{figure}
\includegraphics[width=1\linewidth,trim=0.7cm 0.7cm 0.7cm 0.7cm]{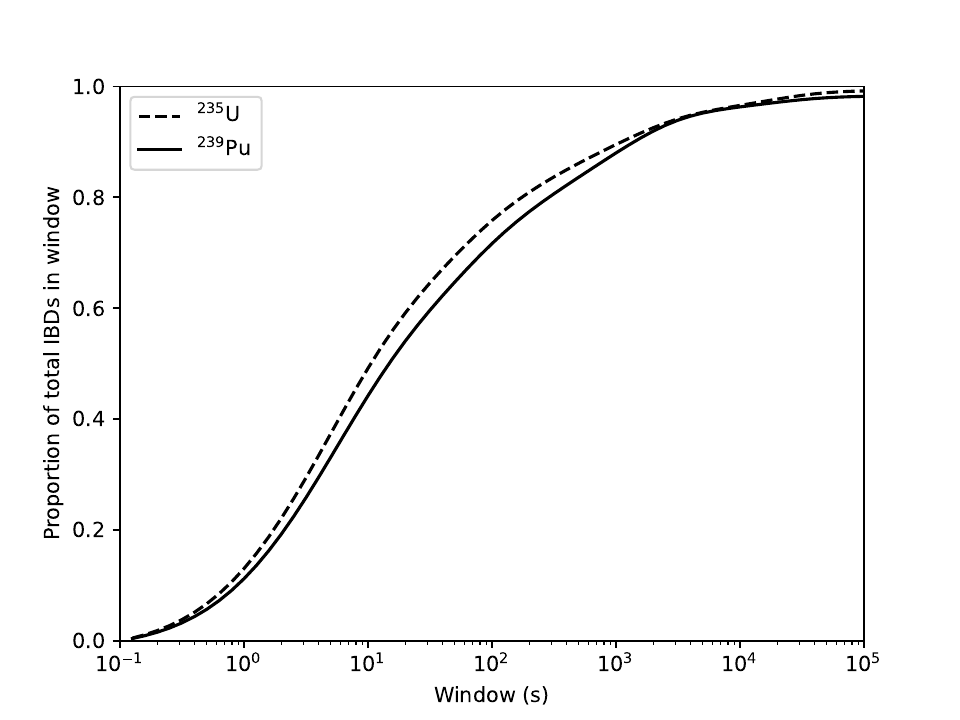}
    
    \caption{Proportion of the total number of inverse beta interactions within a given post-explosion time window for explosions fueled by purely $^{235}$U (dashed) or $^{239}$Pu (solid).}
    \label{fig:IBD_Window}
\end{figure}

The evolution of the neutrino signal with time is an important element of a fission exclusion analysis. Figure \ref{fig:IBD_Window} shows the fraction of IBD interactions expected to occur within a post-explosion time window extending from 0.1~s to just over one day. If collecting the maximum signal were the only consideration, a collection window of about a day would be preferable. However, any realistic IBD detector will register some background events, and the higher background expectation in a long collection window would degrade the sensitivity of the analysis. The ideal window length depends on the background rate and desired sensitivity in a particular monitoring scenario. As a representative example, this analysis sets the collection window length at 10 s. That window is long enough to collect nearly half the IBD signal but short enough to keep background expectations near zero, as discussed in the following sections.

An additional signal consideration, which has only minor impacts on this analysis, is neutrino flavor oscillation. This effect is included in IBD signal rate calculations, using the standard three-flavor neutrino mixing model and global averages for the mixing parameters \cite{ParticleDataGroup:2024cfk}. In this analysis, the impact of flavor oscillations is mainly visible at source-to-detector distances of 10-100~km, where $\theta_{12}$-driven oscillations increase the required detector mass for a given monitoring scenario.

\section{Detector model}
\label{sec:detector}

Beyond the signal parameters described in Sec. \ref{sec:source}, the other key parameters in this analysis belong to the detector: its active mass, the hydrogen content of that active mass, the IBD detection efficiency, and the rate of IBD-mimicking background events. This section describes assumptions made about each of these parameters. The assumptions are kept as generic as possible, and all values are connected to past and current neutrino physics experiments. 

Considering the parameters of detector mass and hydrogen content, it is useful to survey the broad range of technologies that have performed IBD measurements using nuclear reactors as the neutrino source. Liquid scintillator reactor neutrino detectors have been constructed with active masses spanning many orders of magnitude, from the ton or sub-ton scale (PROSPECT~\cite{PROSPECT:2018dnc}, STEREO~\cite{STEREO:2022nzk}, NEOS~\cite{bib:neos}, and others~\cite{Bowden:2006hu,bib:Bugey3,Serebrov:2020kmd,Yang:2025cxa})
through tens or hundreds of tons (the $\theta_{13}$ experiments~\cite{RENO:2012mkc,DayaBay:2015kir,DoubleChooz:2022ukr}) to kilotons (KamLAND \cite{KamLAND:2008dgz}, JUNO \cite{JUNO:2021vlw}). Water-based reactor neutrino detectors have also been built at scales from a few tons (Eos~\cite{Anderson:2022lbb} and others~\cite{Kemp:2024yth}) to a few kilotons (SNO+ water phase~\cite{SNO:2024wzq}). Even larger water detectors may detect reactor neutrinos within this decade (Super-Kamiokande at about 20 kt~\cite{Super-Kamiokande:2002weg,Super-Kamiokande:2021the}, Hyper-Kamiokande at nearly 200 kt~\cite{Hyper-Kamiokande:2018ofw}). Solid plastic scintillator detectors for reactor neutrinos have had masses of about a ton (MAD~\cite{10337889}, Soli$\delta$~\cite{SoLid:2020cen}, DANSS~\cite{DANSS:2018fnn}) or less (miniCHANDLER~\cite{Haghighat:2018mve}, ROADSTR~\cite{Rodrigues:2025hqk}, and others~\cite{Oguri:2014gta, Bridges:2022khb}). For this analysis, all calculations assume a detector target with a hydrogen density of 5.66$\times$10$^{25}$ per kg, similar to that of water (6.76$\times$10$^{25}$ H/kg), pulse-shape-discriminating liquid scintillator (5.76$\times$10$^{25}$ H/kg), pulse-shape-discriminating plastic scintillator (4.26$\times$10$^{25}$ H/kg), and polyvinyl toluene-based plastic scintillator (5.76$\times$10$^{25}$ H/kg)~\cite{eljen_EJ309_2021}. From the examples cited, it should be clear that this assumption is reasonable for a wide range of detector masses.

Achieved IBD detection efficiencies have also varied widely between reactor neutrino experiments due to differences in detector designs and background rejection requirements. While IBD experiments generally rely on detection of a time coincidence between one prompt IBD positron and one delayed IBD neutron capture, they vary considerably in their choice of neutron-capture target (hydrogen or dopants such as gadolinium or lithium) and use of other distinguishing signal characteristics (event topology, directionality, fraction of scintillation versus Cherenkov light, etc.). This analysis assumes an IBD detection efficiency of 90\%. Several of the experiments listed above have demonstrated efficiencies exceeding 50\%. Results of the analysis can naturally be scaled to lower efficiencies, as well as other values of hydrogen density. 

The last detector parameter required for a fission exclusion analysis is the rate of IBD-like background events. In reactor neutrino searches, backgrounds typically arise from ambient radioactivity or cosmic ray interactions. The rate of such events varies widely with detector type, size, and location, including depth underground. As Sec. \ref{sec:sensitivity} will explain, a requirement for the fission exclusion measurement in this paper is a full-detector IBD background rate lower than 5$\times$10$^{-3}$ per second. Reactor neutrino experiments of broadly varying size and technology have demonstrated background rates at or below this level. For underground detectors like Daya Bay, KamLAND, and SNO+, rock overburden greatly reduces cosmogenic IBD backgrounds, while shielding systems, materials radiopurity screening, and targeted analysis cuts reduce both radiogenic and cosmogenic backgrounds.  These experiments achieved full-detector background rates of $\sim$10$^{-3}$~\cite{DayaBay:2012fng}, $<$10$^{-4}$~\cite{KamLAND:2008dgz}, and $<$10$^{-6}$~\cite{SNO:2022qvw}) per second, respectively. In a very different experimental context, the on-surface PROSPECT experiment relied on passive shielding, optical segmentation, and pulse shape discrimination to achieve a full-detector background rate of 3.6$\times$10$^{-3}$~per second ~\cite{PROSPECT:2022wlf}. Considered together, these measurements justify the assumption of backgrounds rates below 5$\times$10$^{-3}$ per second for detector masses up to at least a few kilotons.

All of the detectors mentioned so far in this section are one-of-a-kind instruments designed by science teams to advance neutrino physics. It is worth noting that more readily deployable applications-oriented neutrino detectors may also be able to meet the signal and background standards for a fission exclusion analysis. In the United States, two R\&D campaigns are currently pursuing applications-oriented IBD detectors with different target materials and sensing modalities. The Eos detector is demonstrating hybrid Cherenkov-scintillation light detection and exploring the feasibility of directional event reconstruction with a 4 t target mass. Eos can deploy target materials ranging from water and water-based liquid scintillators to pure scintillators \cite{Anderson_2023}. 
The Mobile Antineutrino Demonstrator (MAD), containing about one ton of segmented plastic scintillator inside a passively shielded mobile enclosure, aims to advance that technology toward in-the-field applications ~\cite{10337889}. 
In the coming years, these two efforts 
will explore the suitability of these technologies for the fission yield exclusion use case, among others. In particular, these efforts can determine what background rates and IBD detection efficiencies are likely to be achieved in realistic detector deployment scenarios.  

\section{Fission Exclusion Calculation}
\label{sec:sensitivity}

Using the signal and detector models in the previous sections, it is possible to calculate the fission yield that can be excluded in a range of scenarios. For this analysis, each scenario has the same basic setup: A neutrino detector with a specified mass is located a specified distance from the chemical explosion (or other event) that is to be monitored. Data is taken in the 10 s immediately following the detonation (or data may be taken over a longer period, with the 10 s window picked out of the full datastream using knowledge of the precise time of the event in question). It is assumed for this analysis that zero IBD candidates are observed in that 10 s window (which amounts to a certain assumption about background levels, as explained below). Based on that null observation, a limit is set on the fission yield that could possibly have accompanied the explosion. The final result reported to observers, regulators, etc. would be a statement of this form: ``Data from the neutrino detector indicates, with 95\% confidence, that there was no more than $Y$ tons of fission yield in the explosion that occurred at (specified location) at (specified time)." The confidence level could be set higher or lower than 95\%. That level is used for the rest of this paper as a representative case.

Choosing a 95\% confidence level for limits means that fission yields \textit{above} the calculated limit would, in at least 95\% of cases, produce one or more signal events in the collection window. This amounts to the following condition: 
\begin{equation}
P(N=0 \mid \lambda_s) = e^{-\lambda_s} < 0.05
\end{equation}
where $N$ is the number of observed signal events and $P$ is the Poisson probability of observing $N$ events given an expectation of $\lambda_s$ events. That condition produces $\lambda_s \geq 3$ events. 

To set a limit in a given scenario, one therefore needs to determine the fission yield that would lead to a mean signal of three events in the collection window. That fission yield $Y$ is determined from the following relationship:
\begin{equation}
\hspace*{-0.23cm}
\lambda_s(Y, d, M) \! = \! N_p(M) \epsilon \! \int \! \phi(E; Y, d)\, P_{ee}(E,d)\, \sigma(E)\, dE.
\end{equation}
Here, 
$d$ is the distance from the detector center to the explosion, 
$M$ is the detector mass, 
$N_p$ is the number of protons in the detector (a function of the mass),
$\epsilon$ is the detector efficiency,
$\phi$ is the neutrino flux,
$E$ is the neutrino energy, 
$P_{ee}$ is the electron flavor survival probability assuming standard neutrino flavor oscillations, and
$\sigma(E)$ is the IBD cross section, 
with the last six parameters handled as discussed in Sec. \ref{sec:source}. 
In particular, the flux is scaled from a CONFLUX simulation of the first 10 s of neutrino emission following a 1000~t yield pure $^{239}$Pu or $^{235}$U explosion. That simulation is scaled linearly to obtain the yield of interest and geometrically with distance according to the inverse-square expectation from a point-like isotropic source:
\begin{equation}
\phi(E; Y, d) \;=\; \phi_{\text{ref}}(E) \; \frac{Y}{Y_{\text{ref}}} \; \frac{1}{4 \pi d^2}
\end{equation}
where $\phi_{ref}(E)$ is the flux for the reference simulation at $Y_{ref} = 1000$ t of yield.

As noted above, the expectation that there will be zero background events in the signal collection window constrains the permissible background rate in each scenario. This analysis assumes that there will be zero background events in the 10 s signal collection window in 95\% of cases. (Like the confidence level and window length, this tolerance could be set lower or higher in a real monitoring scenario.) For the Poisson-distributed backgrounds one would expect in a neutrino detector, the assumption of zero backgrounds in 95\% of cases corresponds to the condition
\begin{equation}
P(N \geq 1 \mid \lambda_b) = 1 - e^{-\lambda_b} < 0.05
\end{equation}
where $N$ is the number of observed background events and $P$ is the Poisson probability of observing $N$ events given an expectation of $\lambda_b$. That condition produces $\lambda_b < 0.05$ events in the 10 s window. This result corresponds to a maximum permissible background rate of $5 \times 10^{-3}$ events per second or 443 events per day. Note that the approach described here is a simple counting analysis that does not take into account the temporal distribution of expected signal and background events within the 10 s window. Using time series information would improve signal-to-background discrimination and thus raise the permissible background rate and/or allow extension of the signal collection window beyond the 10~s assumed here.

At this point, one may wonder whether the monitored explosion could create background events on top of the Poisson-distributed expectation from ambient radioactivity and cosmic ray interactions. Certainly, a fission explosion would produce a neutron flux that could increase the IBD background rate, but a chemical explosion would not. A very nearby chemical explosion could induce electromagnetic signals in a neutrino detector that could transiently raise IBD backgrounds, a possibility that would require more investigation before a real deployment. However, such transients would likely fade quickly enough that a background-free neutrino measurement could still be attempted by opening the signal collection window slightly after the detonation.

\section{Physically Inaccessible Regions}
\label{sec:exclusion}

\begin{figure*}[ht]
    \centering
    
    \includegraphics[width=0.9\linewidth,trim=0.8cm 0.6cm 0.8cm 0.8cm]{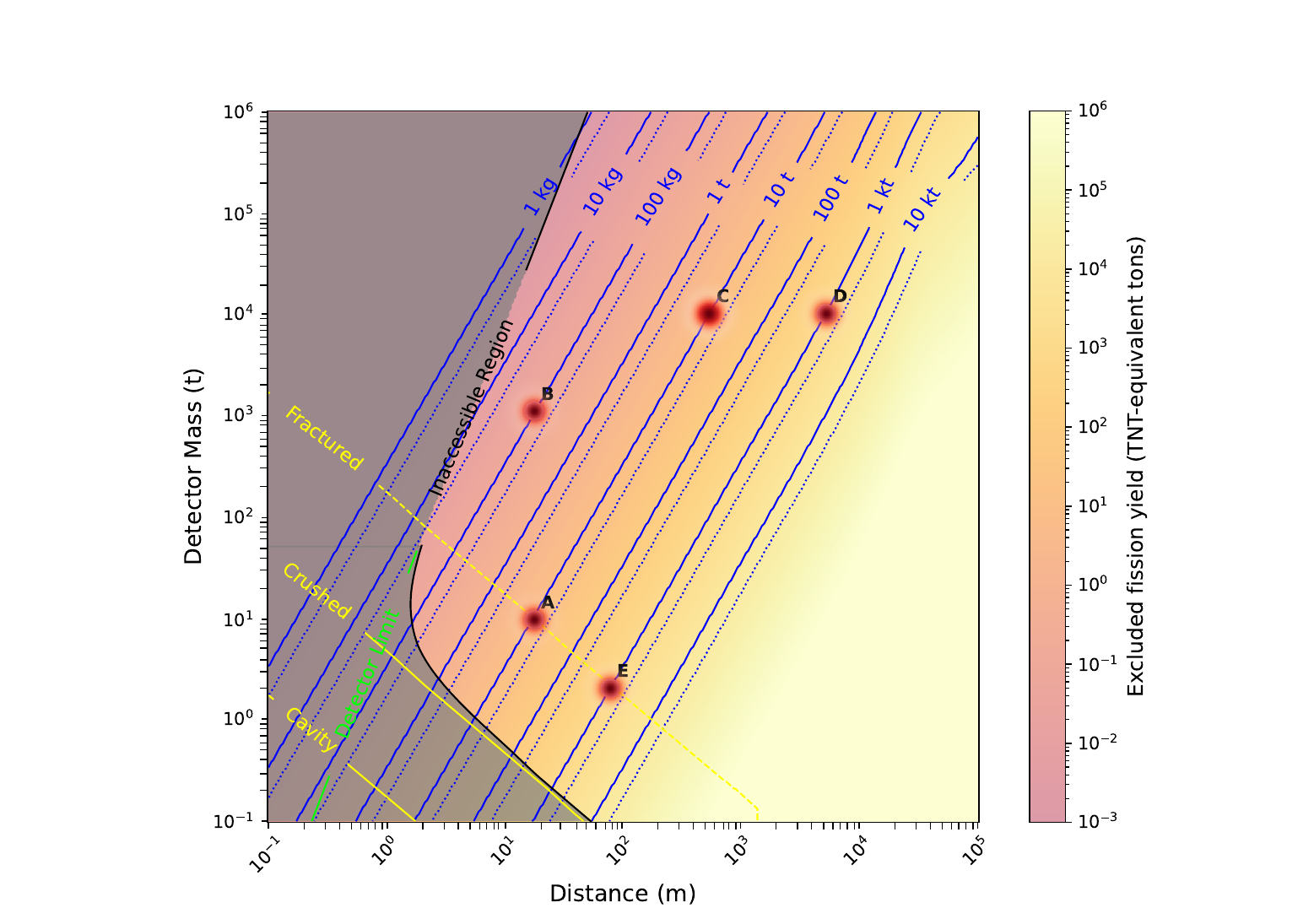}
       \caption{Fission yield that can be excluded in a monitored explosion, at 95\% confidence level, as a function of neutrino detector mass and distance between the explosion and detector center. Solid blue curves (and yellow-to-pink background shading) mark upper limits on fission yield that could have occurred in a $^{239}$Pu explosion, and dotted blue curves mark the equivalent for $^{235}$U. Yellow lines mark the outer extent of damage zones from an underground explosion, as explained in the text. (Corresponding yellow contours appear in Fig.~\ref{fig:blast_regions}.) The green detector limit line marks the spatial extent of the detector. Combining the solid yellow and green contours gives the inaccessible region, shaded gray, in which detector placement is not possible. The highlighted points A-E indicate a sample of potential operating regimes.}
    \label{fig:sens}
\end{figure*}

Using the calculation described in the previous section, it is possible to construct a plot of excludable fission yield as a function of detector mass and distance from the monitored explosion. Figure \ref{fig:sens} shows that result. Importantly, some regions of the plot (shaded gray) are not physically accessible. That is, there are certain combinations of standoff distance and detector mass that cannot be physically realized. The basic reason is that the monitored explosion and its blast effects must occur entirely outside the detector, at least in the basic single-detector setup envisioned in this paper. More precisely, the inaccessible region arises from two effects, one related to the spatial extent of the detector and one related to the spatial extent of the explosion. The exact size of these effects depends on details of the explosion, detector geometry, and other features of the specific deployment scenario. However, a general idea of the boundaries can be derived from the considerations described below.

Since this analysis assumes a detector of a fixed density (roughly that of organic scintillator or water), the spatial dimensions of the detector will scale with detector mass. For simplicity, this analysis assumes a cubic detector shape. The distance from the center to side of the active mass scale, called $R_{det}$, is then
\begin{equation}
R_{det} = \frac{(M/\rho)^{1/3}}{2}.
\end{equation}
where $M$ is the active mass of the detector and $\rho$ is its mass density as specified in Sec. \ref{sec:detector}. To occur outside the active detector mass, the explosion would certainly need be located at a distance greater than $R_{det}$ from the center of the detector. Practically speaking, to be located fully outside the detector, the explosion would likely need to be located even farther out, since a real neutrino detector would include layers of material outside the active mass (photomultiplier tubes or other light sensors, shielding, electronics, and perhaps a buffer volume). The minimal requirement of occurring outside the active mass appears as the green detector limit line in Fig. \ref{fig:sens}. As noted, this limit assumes a single cubic detector. Alternative setups involving specialized geometries or multi-detector arrays might be considered, but on the logarithmic scale of Fig. \ref{fig:sens}, their detector limits would not be dramatically different.

Requiring that the detonation of the explosion occur outside the detector is, of course, not the only practical constraint in the case of a large explosion. There must be enough material between the detonation point and detector to absorb the blast effects and protect the detector from incapacitating damage. The spatial extent of damage from an explosion depends on the local environment, but some basic scaling relationships apply to explosions of both chemical and nuclear origin conducted in underground rock \cite{Stroujkova:2016}. The large chemical explosions conducted and planned by NNSA on the former Nevada test site fall in this category. Explosions for subcritical experiments are fully enclosed within rigid containment structures, so a rock model may overestimate the extent of damage in these cases but should provide a conservative bound for deployment constraints.

\begin{figure}[t]
    \centering
    \resizebox{\columnwidth}{!}{
    \begin{tikzpicture}
        \fill[brown] (0,0) rectangle (10,5);

        \def\Y{3}
        \def\Ycube{pow(\Y,1/3)}

        \def\cx{2.5}
        \def\cy{2.5}

        \def\rcavity{8}
        \def\rcrushed{20}
        \def\rfractured{30}

        \filldraw[fill=brown, draw=yellow, dashed, line width=2pt] (\cx,\cy) circle ({\rfractured*\Ycube/20});
        \filldraw[fill=gray, draw=yellow, solid, line width=2pt]  (\cx,\cy) circle ({\rcrushed*\Ycube/20});
        \filldraw[fill=gray, draw=yellow, solid, line width=2pt](\cx,\cy) circle ({\rcavity*\Ycube/20});

        \def\vx{7}
        \def\sA{2}
        \def\vy{\cy - \sA/2} 

        \definecolor{mygreen}{RGB}{102,205,100}
        \fill[white] (\vx,\vy) rectangle ({\vx+\sA},{\vy+\sA});

        \draw[black, line width=2pt] 
            (\cx,\cy) -- ({\vx+\sA/2},{\vy+\sA/2}) 
            node[midway, above, sloped, font=\small, inner sep=3pt] {\; \; Distance};
             
        \node[anchor=center, white, font=\small] at ({\vx+\sA/2},\vy+\sA+.2) {Detector};

        \node[anchor=center, yellow, font=\scriptsize] at (\cx,\cy+.2) {Cavity};
        \node[anchor=center, yellow, font=\scriptsize] at ({\cx},\cy+.8) {Crushed rock};
        \node[anchor=center, yellow, font=\scriptsize] at ({\cx},\cy+1.7) {Fractured rock};
    \end{tikzpicture}
    }
    \caption{Schematic diagram showing the three explosion damage zones defined in this analysis as well as the source-to-detector distance, defined to run from the detonation at the cavity center to the detector center. As described in the text, no part of the detector may be located within the cavity or crushed rock region, and deployments within the fractured rock region would bring engineering challenges.}
    \label{fig:blast_regions}
\end{figure}
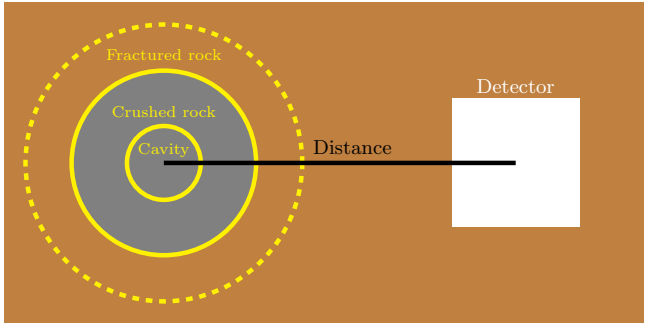

In underground explosions, three concentric zones of damage are commonly identified: a cavity, centered on the detonation, in which the surrounding rock is either vaporized or pushed out from the blast wave; a layer of ``crushed'' rock, in which the rock is permanently damaged; and a layer of ``fractured'' rock, in which the rock suffers elastic strain and smaller cracks. These zones are illustrated in Fig.~\ref{fig:blast_regions}. Decades of observations of both chemical and nuclear explosions have indicated that the radius of all three regions tends to scale with the cube root of the explosion yield:
\begin{equation}
R_{exp} = k \, \left( \frac{Y}{1 \, \text{kt}} \right)^{1/3}.
\end{equation}
Here, $R_{exp}$ is the estimated radial extent of each zone and $k$ is a scaling coefficient. This analysis uses approximate values of $k$ derived from observations \cite{Stroujkova:2016, Rogers:1970, Glasstone:1977}. For the cavity, $k=8~\text{m}$; for the crushed region, $k=24~\text{m}$; and for the fractured region, $k=80~\text{m}$. This analysis assumes that no part of the detector may be located within the cavity or crushed rock region, because the damage in these regions would almost certainly render a neutrino detector inoperable. Summing constraints from the detector size and the crushed region boundary defines the inaccessible region in Fig. \ref{fig:sens}. Deployment of a neutrino detector in the fractured rock region is not entirely inconceivable, but engineering requirements for such a deployment would need to be carefully considered.

\section{Results}
\label{sec:results}

Figure~\ref{fig:sens} shows the main result of this analysis: the fission yield that can be excluded by a neutrino detector of a given mass at a given standoff distance, following the assumptions in Secs. \ref{sec:source}-\ref{sec:sensitivity}. Since both axes of Fig. \ref{fig:sens} have logarithmic scales, the major features of the plot are fairly insensitive to moderate shifts in these assumptions. A prominent element of Fig. \ref{fig:sens} is the region of physically inaccessible deployment scenarios described in Sec. \ref{sec:exclusion}. To connect the figure to possible uses on former nuclear test sites, it is useful to make some observations about both the accessible and inaccessible regions.

An important feature of the inaccessible region is that it almost completely covers the 1 kg yield contour, at least for the conservative assumption (see Sec. \ref{sec:source}) of a $^{239}$Pu source. This means that a fission yield limit of 1~kg or less is not achievable in a neutrino detector with less than about 500 kt of active mass. For comparison, the largest IBD-sensitive neutrino detector currently under construction, Hyper-Kamiokande, will have an active mass of less than 200 kt \cite{Hyper-Kamiokande:2018ofw}. The practical inaccessibility of the 1 kg yield line has a major consequence for the idea of monitoring subcritical experiments, one of the activities of significant interest on former nuclear test sites. The fission yield threshold that separates subcritical from supercritical experiments is far below 1 kg. Open-source estimates have placed it at less than one gram for a variety of possible test configurations \cite{JonesVonHippel1997}. Even if one is satisfied with the less stringent goal of distinguishing subcritical experiments from so-called hydronuclear tests, the yield threshold is still around the kilogram level \cite{JonesVonHippel1997}. Therefore, this analysis indicates that neutrino detectors are unlikely to have practical value in excluding anomalous fission yield in subcritical experiments.

In contrast, Fig. \ref{fig:sens} shows a range of possibilities for excluding fission yield in large chemical explosions, such as those at the NNSS. Within the accessible region are many combinations of detector size, standoff distance, and excludable fission yield that could be relevant to monitoring these events, which may involve up to 10 t of chemical explosive yield. The five red points labeled A-E in the figure highlight a sample of possible operating regimes:

\begin{description}

\item[A]  It is possible to demonstrate that a 10 t yield explosion is not entirely fission using a neutrino detector with a mass of about 10 t located $\sim$20~m from the source. 

\item[B] Alternatively, one can demonstrate that a 10 t yield explosion was no more than 1\% fission yield (i.e., contained no more than 100 kg fission yield) using a neutrino detector with a few kilotons of mass at a distance of about 10 m. 

\item[C] A single detector with a mass of about 10 kt could
demonstrate that a 10 t yield explosion located anywhere within hundreds of meters was not entirely fission. This capability would allows coverage of multiple explosions with a single detector.

\item[D] The not-entirely-fission nature of a 1 kt yield explosion (scale of the largest chemical explosions ever conducted at NNSS \cite{NPE:1994}) could be confirmed with a neutrino detector of mass about 10 kt at a distance of a few km. 

\item[E] Equivalently, the not-entirely-fission nature of a 1 kt yield explosion could be confirmed with a neutrino detector of mass about 2 t placed as far as 100 m from the explosion. 

\end{description}

The accessible region extends beyond the boundaries of Fig. \ref{fig:sens}, but points beyond those bounds correspond to scenarios of less practical interest. For example, verifying the non-fission nature of explosions on the 10 t scale over a region that covers $\sim$100~km standoffs (approximately the scale of the entire NNSS) using a single detector would require that the detector mass exceed 1~Mt. Such a concept appears impractical, in agreement with a previous study centered on longer-range fission explosion monitoring \cite{Foxe:2020}.

\section{Conclusions}
\label{sec:conclusions}

This paper has analyzed the potential for neutrino detectors to exclude fission yield in relatively nearby events, such as large chemical explosions and subcritical experiments occurring on former nuclear test sites. It is the first such study to include a range of source strengths and an empirically grounded treatment of backgrounds. 
For certain source strengths and baselines, inverse beta decay detectors could set potentially useful limits on fission yield. For example, a detector on the 10 t mass scale, placed about 10 m from a planned chemical explosion with 10 t of yield, could confirm that the explosion contained no more than 10\% fission yield. A detector with a mass around 1000 t located at the same place could set a tighter limit of 1\% fission yield.
Other source strengths and baselines present more constrained scenarios.
Consistent with previous studies of long-range neutrino explosion monitoring, this analysis finds that excluding fission yield in relatively distant events ($\gtrsim 100$ km) requires impractically large inverse beta decay detectors ($\gtrsim 1$ Mt), even for the relatively large excluded fission yield of 1 kt. Inverse beta decay detectors on a similarly impractical scale would be required to confirm that a subcritical experiment has not reached critical levels of fission yield, even when monitoring occurs at the closest possible baseline. 
These calculations provide a quantitative basis for considering whether neutrino detectors could meet the policy goals and implementation constraints of particular explosion monitoring scenarios.

\begin{acknowledgments}

This work was performed under the auspices of the U.S. Department of Energy by: 
Brookhaven National Laboratory under Contract DE-AC02-98CH10886;
Lawrence Berkeley National Laboratory under Contract DE-AC02-05CH11231; 
Lawrence Livermore National Laboratory, Livermore, CA, USA under Contract DE-AC52-07NA27344; 
UT-Battelle, LLC, under contract number DE-AC05-00OR22725; and
Pacific Northwest National Laboratory under Contract DE-AC05-76RL01830.
The project was funded by the U.S. Department of Energy, National Nuclear Security Administration, Office of Defense Nuclear Nonproliferation  Research and Development (DNN R\&D).
We further acknowledge support from the National Institute of Standards and Technology. 

The U.S. government retains and the publisher, by accepting the article for publication, acknowledges that the U.S. government retains a nonexclusive, paid-up, irrevocable, worldwide license to publish or reproduce the published form of this manuscript, or allow others to do so, for U.S. government purposes. DOE will provide public access to these results of federally sponsored research in accordance with the DOE Public Access Plan (https://www.energy.gov/doe-public-access-plan).

The views expressed in this paper are those of the authors and do not reflect the official policy or position of the U.S. Naval Academy, Department of the Navy, Department of Defense, or U.S. Government. 

LLNL-JRNL-2019606. 

\end{acknowledgments}

\bibliography{testsite}

@article{KamLAND:2008dgz,
    author = "Abe, S. and others",
    collaboration = "KamLAND",
    title = "{Precision Measurement of Neutrino Oscillation Parameters with KamLAND}",
    eprint = "0801.4589",
    archivePrefix = "arXiv",
    primaryClass = "hep-ex",
    doi = "10.1103/PhysRevLett.100.221803",
    journal = "Phys. Rev. Lett.",
    volume = "100",
    pages = "221803",
    year = "2008"
}

@article{SNO:2024wzq,
    author = "Allega, A. and others",
    collaboration = "SNO+",
    title = "{Initial measurement of reactor antineutrino oscillation at SNO+}",
    eprint = "2405.19700",
    archivePrefix = "arXiv",
    primaryClass = "hep-ex",
    doi = "10.1140/epjc/s10052-024-13687-5",
    journal = "Eur. Phys. J. C",
    volume = "85",
    number = "1",
    pages = "17",
    year = "2025",
    note = "[Erratum: Eur.Phys.J.C 85, 296 (2025)]"
}

@article{PROSPECT:2022wlf,
    author = "Andriamirado, M. and others",
    collaboration = "(PROSPECT)",
    title = "{Final Measurement of the U235 Antineutrino Energy Spectrum with the PROSPECT-I Detector at HFIR}",
    eprint = "2212.10669",
    archivePrefix = "arXiv",
    primaryClass = "nucl-ex",
    doi = "10.1103/PhysRevLett.131.021802",
    journal = "Phys. Rev. Lett.",
    volume = "131",
    number = "2",
    pages = "021802",
    year = "2023"
}

@article{SNO:2022qvw,
    author = "Allega, A. and others",
    collaboration = "SNO+",
    title = "{Evidence of Antineutrinos from Distant Reactors using Pure Water at SNO+}",
    eprint = "2210.14154",
    archivePrefix = "arXiv",
    primaryClass = "nucl-ex",
    doi = "10.1103/PhysRevLett.130.091801",
    journal = "Phys. Rev. Lett.",
    volume = "130",
    number = "9",
    pages = "091801",
    year = "2023"
}

@article{JUNO:2021vlw,
    author = "Abusleme, Angel and others",
    collaboration = "JUNO",
    title = "{JUNO physics and detector}",
    eprint = "2104.02565",
    archivePrefix = "arXiv",
    primaryClass = "hep-ex",
    doi = "10.1016/j.ppnp.2021.103927",
    journal = "Prog. Part. Nucl. Phys.",
    volume = "123",
    pages = "103927",
    year = "2022"
}

@article{Haghighat:2018mve,
    author = "Haghighat, Alireza and Huber, Patrick and Li, Shengchao and Link, Jonathan M. and Mariani, Camillo and Park, Jaewon and Subedi, Tulasi",
    title = "{Observation of Reactor Antineutrinos with a Rapidly-Deployable Surface-Level Detector}",
    eprint = "1812.02163",
    archivePrefix = "arXiv",
    primaryClass = "physics.ins-det",
    doi = "10.1103/PhysRevApplied.13.034028",
    journal = "Phys. Rev. Applied",
    volume = "13",
    number = "3",
    pages = "034028",
    year = "2020"
}

@article{DayaBay:2012fng,
    author = "An, F. P. and others",
    collaboration = "Daya Bay",
    title = "{Observation of electron-antineutrino disappearance at Daya Bay}",
    eprint = "1203.1669",
    archivePrefix = "arXiv",
    primaryClass = "hep-ex",
    doi = "10.1103/PhysRevLett.108.171803",
    journal = "Phys. Rev. Lett.",
    volume = "108",
    pages = "171803",
    year = "2012"
}

@misc{Reines,
    author  = "Frederick Reines",
    title   = "The {Neutrino: From Poltergeist to Particle}",
    year    = "Nobel Lecture, 1995",
}

@article{LANL1980s,
    author  = "Herald W. Kruse and Rosalie Loncoski and Joseph M. Mack",
    title   = "Antineutrino detector for $\nu$ oscillation studies at fission weapon tests and at {LAMPF}",
    year    = "1981",
    journal = "IEEE",
}

@article{annrev,
title = {Concepts for Neutrino Applications},
author = {Akindele, Oluwatomi A. and Carr, Rachel},
doi = {10.1146/annurev-nucl-102122-023751},
journal = {Annual Review of Nuclear and Particle Science},
number = 1,
volume = 74,
place = {United States},
year = {2024},
month = {9}
}

@article{Bernstein2001,
    author  = "Adam Bernstein and Todd West and Vipin Gupta",
    title   = "An Assessment of Antineutrino Detection as a Tool for Monitoring Nuclear Explosions",
    year    = "2001",
    journal = "Science and Global Security",
}

@article{Carr2017,
    author = "Carr, Rachel and Dalnoki-Veress, Ferenc and Bernstein, Adam",
    title = "{Sensitivity of seismically-cued antineutrino detectors to nuclear explosions}",
    eprint = "1712.04001",
    archivePrefix = "arXiv",
    primaryClass = "nucl-ex",
    doi = "10.1103/PhysRevApplied.10.024014",
    journal = "Phys. Rev. Applied",
    volume = "10",
    number = "2",
    pages = "024014",
    year = "2018"
}

@misc{nnsa_press_release_2024,
  author       = {{National Nuclear Security Administration}},
  title        = "{NNSA demonstrates transparency in its operations and support for the Comprehensive Nuclear-Test-Ban Treaty by hosting visitors at Sandia and Nevada site}",
  howpublished = {Press release},
  institution  = {U.S. Department of Energy},
  year         = {2024},
  month        = {June},
  url          = {https://www.energy.gov/nnsa/articles/nnsa-demonstrates-transparency-its-operations-and-support-comprehensive-nuclear-test},
  note         = {Accessed: 2024-07-15},
}

@article{reuters2023,
  author = {Andrew Osborn},
  title = "{Russia accuses US of nuclear testing site activity, says it won't test unless US does}",
  journal = "Reuters",
  year = 2023,
  url = {https://www.reuters.com/world/russia-would-only-resume-nuclear-testing-when-us-does-agencies-2023-10-10/},
  urldate = {2023-10-10}
}

@article{cnn2023,
  author = {Eric Cheung, Brad Lendon and Ivan Watson},
  title = "{Exclusive: Satellite images show increased activity at nuclear test sites in Russia, China and US}",
  year = 2023,
  journal = "CNN",
  url = {https://www.cnn.com/2023/09/22/asia/nuclear-testing-china-russia-us-exclusive-intl-hnk-ml/index.html},
  urldate = {2023-09-23}
}

@article{Zhang:2025ayc,
    author = "Zhang, Xianyi and Irani, Anosh and Mendenhall, Michael P. and Rybicki, Nathan and Hayen, Leendert and Bowden, Nathaniel and Huber, Patrick and Littlejohn, Bryce and Bogetic, Sandra",
    title = "{CONFLUX: A standardized framework to calculate reactor antineutrino flux}",
    eprint = "2503.18966",
    archivePrefix = "arXiv",
    primaryClass = "physics.ins-det",
    reportNumber = "LLNL-JRNL-872396",
    doi = "10.1016/j.cpc.2025.109831",
    journal = "Comput. Phys. Commun.",
    volume = "317",
    pages = "109831",
    year = "2025"
}

@article{Vogel:1999zy,
      author         = "Vogel, P. and Beacom, John F.",
      title          = "{Angular distribution of neutron inverse beta decay,
                        anti-neutrino(e) + p $\rightarrow$ $e^+$ + n}",
      journal        = "Phys. Rev.",
      volume         = "D60",
      year           = "1999",
      pages          = "053003",
      doi            = "10.1103/PhysRevD.60.053003",
      eprint         = "hep-ph/9903554",
      archivePrefix  = "arXiv",
      primaryClass   = "hep-ph",
      SLACcitation   = "%%CITATION = HEP-PH/9903554;%%"
}

@article{ParticleDataGroup:2024cfk,
    author = "Navas, S. and others",
    collaboration = "Particle Data Group",
    title = "{Review of particle physics}",
    doi = "10.1103/PhysRevD.110.030001",
    journal = "Phys. Rev. D",
    volume = "110",
    number = "3",
    pages = "030001",
    year = "2024"
}

@article{Kemp:2024yth,
    author = "Kemp, Ernesto and Vieira dos Santos, Willian and dos Anjos, Jo{\~a}o C. and Chimenti, Pietro and Gomez Gonzalez, Luis Fernando and Guedes, Germano P. and Lima, Herman P. and Antunes N{\'o}brega, Rafael and Pepe, Iuri Muniz and Barbosa dos Santos Ribeiro, Dion",
    title = "{Results from ON-OFF Analysis of the Neutrinos-Angra Detector}",
    eprint = "2407.20397",
    archivePrefix = "arXiv",
    primaryClass = "hep-ex",
    doi = "10.1007/s13538-024-01667-9",
    journal = "Braz. J. Phys.",
    volume = "55",
    number = "1",
    pages = "39",
    year = "2025"
}

@article{Anderson:2022lbb,
    author = "Anderson, T. and others",
    title = "{Eos: conceptual design for a demonstrator of hybrid optical detector technology}",
    eprint = "2211.11969",
    archivePrefix = "arXiv",
    primaryClass = "physics.ins-det",
    doi = "10.1088/1748-0221/18/02/P02009",
    journal = "JINST",
    volume = "18",
    number = "02",
    pages = "P02009",
    year = "2023"
}

@article{PROSPECT:2018dnc,
    author = "Ashenfelter, J. and others",
    collaboration = "PROSPECT",
    title = "{The PROSPECT Reactor Antineutrino Experiment}",
    eprint = "1808.00097",
    archivePrefix = "arXiv",
    primaryClass = "physics.ins-det",
    doi = "10.1016/j.nima.2018.12.079",
    journal = "Nucl. Instrum. Meth. A",
    volume = "922",
    pages = "287--309",
    year = "2019"
}

@article{bib:neos,
    author = "Ko, Y.J. and others",
    collaboration = "NEOS",
    title = "{Sterile Neutrino Search at the NEOS Experiment}",
    eprint = "1610.05134",
    archivePrefix = "arXiv",
    primaryClass = "hep-ex",
    doi = "10.1103/PhysRevLett.118.121802",
    journal = "Phys. Rev. Lett.",
    volume = "118",
    number = "12",
    pages = "121802",
    year = "2017"
}

@article{bib:Bugey3,
    author = "Achkar, B. and others",
    title = "{Comparison of anti-neutrino reactor spectrum models with the Bugey-3 measurements}",
    reportNumber = "LAPP-EXP-96-02, CPPM-96-02, ISN-96-10, LPC-96-05",
    doi = "10.1016/0370-2693(96)00216-X",
    journal = "Phys. Lett. B",
    volume = "374",
    pages = "243--248",
    year = "1996"
}

@article{STEREO:2022nzk,
    author = "Almaz\'an, H. and others",
    collaboration = "STEREO",
    title = "{STEREO neutrino spectrum of $^{235}$U fission rejects sterile neutrino hypothesis}",
    eprint = "2210.07664",
    archivePrefix = "arXiv",
    primaryClass = "hep-ex",
    doi = "10.1038/s41586-022-05568-2",
    journal = "Nature",
    volume = "613",
    number = "7943",
    pages = "257--261",
    year = "2023"
}

@article{Serebrov:2020kmd,
    author = "Serebrov, A. P. and others",
    title = "{Search for sterile neutrinos with the Neutrino-4 experiment and measurement results}",
    eprint = "2005.05301",
    archivePrefix = "arXiv",
    primaryClass = "hep-ex",
    doi = "10.1103/PhysRevD.104.032003",
    journal = "Phys. Rev. D",
    volume = "104",
    number = "3",
    pages = "032003",
    year = "2021"
}

@article{Bowden:2006hu,
    author = "Bowden, N. S. and others",
    title = "{Experimental results from an antineutrino detector for cooperative monitoring of nuclear reactors}",
    eprint = "physics/0612152",
    archivePrefix = "arXiv",
    doi = "10.1016/j.nima.2006.12.015",
    journal = "Nucl. Instrum. Meth. A",
    volume = "572",
    pages = "985--998",
    year = "2007"
}

@article{DayaBay:2015kir,
    author = "An, F. P. and others",
    collaboration = "Daya Bay",
    title = "{The Detector System of The Daya Bay Reactor Neutrino Experiment}",
    eprint = "1508.03943",
    archivePrefix = "arXiv",
    primaryClass = "physics.ins-det",
    doi = "10.1016/j.nima.2015.11.144",
    journal = "Nucl. Instrum. Meth. A",
    volume = "811",
    pages = "133--161",
    year = "2016"
}

@article{DoubleChooz:2022ukr,
    author = "de Kerret, H. and others",
    collaboration = "Double Chooz",
    title = "{The Double Chooz antineutrino detectors}",
    eprint = "2201.13285",
    archivePrefix = "arXiv",
    primaryClass = "physics.ins-det",
    doi = "10.1140/epjc/s10052-022-10726-x",
    journal = "Eur. Phys. J. C",
    volume = "82",
    number = "9",
    pages = "804",
    year = "2022"
}

@article{Super-Kamiokande:2021the,
    author = "Abe, K. and others",
    collaboration = "Super-Kamiokande",
    title = "{First gadolinium loading to Super-Kamiokande}",
    eprint = "2109.00360",
    archivePrefix = "arXiv",
    primaryClass = "physics.ins-det",
    doi = "10.1016/j.nima.2021.166248",
    journal = "Nucl. Instrum. Meth. A",
    volume = "1027",
    pages = "166248",
    year = "2022"
}

@article{RENO:2012mkc,
    author = "Ahn, J. K. and others",
    collaboration = "RENO",
    title = "{Observation of Reactor Electron Antineutrino Disappearance in the RENO Experiment}",
    eprint = "1204.0626",
    archivePrefix = "arXiv",
    primaryClass = "hep-ex",
    doi = "10.1103/PhysRevLett.108.191802",
    journal = "Phys. Rev. Lett.",
    volume = "108",
    pages = "191802",
    year = "2012"
}

@misc{Hyper-Kamiokande:2018ofw,
    author = "Abe, K. and others",
    collaboration = "Hyper-Kamiokande",
    title = "{Hyper-Kamiokande Design Report}",
    eprint = "1805.04163",
    archivePrefix = "arXiv",
    primaryClass = "physics.ins-det",
    month = "5",
    year = "2018"
}

@article{Super-Kamiokande:2002weg,
    author = "Fukuda, Y. and others",
    editor = "Ilyin, V. A. and Korenkov, V. V. and Perret-Gallix, D.",
    collaboration = "Super-Kamiokande",
    title = "{The Super-Kamiokande detector}",
    doi = "10.1016/S0168-9002(03)00425-X",
    journal = "Nucl. Instrum. Meth. A",
    volume = "501",
    pages = "418--462",
    year = "2003"
}

@article{SoLid:2020cen,
    author = "Abreu, Y. and others",
    collaboration = "SoLid",
    title = "{SoLid: a short baseline reactor neutrino experiment}",
    eprint = "2002.05914",
    archivePrefix = "arXiv",
    primaryClass = "physics.ins-det",
    doi = "10.1088/1748-0221/16/02/P02025",
    journal = "JINST",
    volume = "16",
    number = "02",
    pages = "P02025",
    year = "2021"
}

@article{DANSS:2018fnn,
    author = "Alekseev, I and others",
    collaboration = "DANSS",
    title = "{Search for sterile neutrinos at the DANSS experiment}",
    eprint = "1804.04046",
    archivePrefix = "arXiv",
    primaryClass = "hep-ex",
    doi = "10.1016/j.physletb.2018.10.038",
    journal = "Phys. Lett. B",
    volume = "787",
    pages = "56--63",
    year = "2018"
}

@article{Bridges:2022khb,
    author = "Bridges, K. and others",
    title = "{VIDARR: monitoring reactor anti-neutrinos using a plastic scintillator detector in a mobile laboratory}",
    doi = "10.1088/1748-0221/17/10/P10009",
    journal = "JINST",
    volume = "17",
    number = "10",
    pages = "P10009",
    year = "2022"
}

@article{Oguri:2014gta,
    author = "Oguri, S. and Kuroda, Y. and Kato, Y. and Nakata, R. and Inoue, Y. and Ito, C. and Minowa, M.",
    title = "{Reactor antineutrino monitoring with a plastic scintillator array as a new safeguards method}",
    eprint = "1404.7309",
    archivePrefix = "arXiv",
    primaryClass = "physics.ins-det",
    doi = "10.1016/j.nima.2014.04.065",
    journal = "Nucl. Instrum. Meth. A",
    volume = "757",
    pages = "33--39",
    year = "2014"
}

@misc{Yang:2025cxa,
    author = "Yang, Byeongsu and others",
    title = "{RENE experiment for the sterile neutrino search using reactor neutrinos}",
    eprint = "2507.22376",
    archivePrefix = "arXiv",
    primaryClass = "hep-ex",
    month = "7",
    year = "2025"
}

@article{Rodrigues:2025hqk,
    author = "Benevides Rodrigues, O. and others",
    title = "{Prototype reactor-antineutrino detector based on 6Li-doped pulse-shape-discriminating plastic scintillator}",
    eprint = "2505.05696",
    archivePrefix = "arXiv",
    primaryClass = "physics.ins-det",
    reportNumber = "LLNL-JRNL-2005361",
    doi = "10.1103/6t97-9bpb",
    journal = "Phys. Rev. Applied",
    volume = "24",
    number = "5",
    pages = "054023",
    year = "2025"
}

@INPROCEEDINGS{10337889,
  author={Ross, J.},
  booktitle={2023 IEEE Nuclear Science Symposium, Medical Imaging Conference and International Symposium on Room-Temperature Semiconductor Detectors (NSS MIC RTSD)}, 
  title={The Mobile Antineutrino Demonstrator Project}, 
  year={2023},
  volume={},
  number={},
  pages={1-1},
  doi={10.1109/NSSMICRTSD49126.2023.10337889}}

@techreport{eljen_EJ309_2021,
  title        = {Neutron/Gamma PSD Liquid Scintillator EJ-301, EJ-309: Product Data Sheet EJ-309},
  author       = {{Eljen Technology}},
  institution  = {Eljen Technology},
  address      = {1300 W. Broadway, Sweetwater, TX 79556, USA},
  year         = {2021},
  type         = {Product Data Sheet},
  number       = {WJ-309},
  url          = {https://eljentechnology.com/products/liquid-scintillators/ej-301-ej-309},
  note         = {Revision Date: Jul 2021},
}

@article{Bernstein:2019hix,
    author = "Bernstein, Adam and Bowden, Nathaniel and Goldblum, Bethany L. and Huber, Patrick and Jovanovic, Igor and Mattingly, John",
    title = "{$Colloquium$: Neutrino detectors as tools for nuclear security}",
    eprint = "1908.07113",
    archivePrefix = "arXiv",
    primaryClass = "physics.soc-ph",
    doi = "10.1103/RevModPhys.92.011003",
    journal = "Rev. Mod. Phys.",
    volume = "92",
    pages = "011003",
    year = "2020"
}

@techreport{PE1:2024,
  author       = {Myers, S. C. and Abbott, G. and Alexander, T. and Alger, E. and Alvarez, A. and Annabelle, N. and Antoun, T. and Auld, G. and Malach, A. and Banuelos, H. and others},
  title        = {A multi-Physics Experiment for Low-Yield Nuclear Explosion Monitoring},
  institution  = {Lawrence Livermore National Laboratory (LLNL), Livermore, CA (United States)},
  doi          = {10.2172/2345984},
  url          = {https://www.osti.gov/biblio/2345984},
  place        = {United States},
  year         = {2024},
  month        = {05}
}

@conference{NPE:1994,
  author       = {Hannon, W J},
  title        = {The non-proliferation experiment},
  booktitle    = {The non-proliferation experiment},
  url          = {https://www.osti.gov/biblio/95982},
  place        = {United States},
  organization = {Lawrence Livermore National Lab., CA (United States)},
  year         = {1994},
  month        = {12}
}

@techreport{SPE:2022,
  author       = {Snelson, Catherine M. and Bradley, Christopher R. and Walter, William R. and Antoun, Tarabay H. and Abbott, Robert A. and Jones, Kyle and Chipman, Veraun D. and Montoya, Lloyd},
  title        = {The Source Physics Experiment (SPE) Science Plan},
  institution  = {Lawrence Livermore National Laboratory (LLNL), Livermore, CA (United States)},
  doi          = {10.2172/1887003},
  url          = {https://www.osti.gov/biblio/1887003},
  place        = {United States},
  year         = {2022},
  month        = {09}
}

@article{Foxe:2020,
  author    = {Foxe, Michael and Bowyer, Theodore and Carr, Rachel and Orrell, John and VanDevender, Brent},
  title     = {Antineutrino Detectors Remain Impractical for Nuclear Explosion Monitoring},
  journal   = {Pure and Applied Geophysics},
  year      = {2020},
  month     = {August},
  volume    = {178},
  number    = {8},
  pages     = {2753--2763},
  publisher = {Springer International Publishing},
  doi       = {10.1007/s00024-020-02464-6},
  url       = {https://doi.org/10.1007/s00024-020-02464-6}
}

@book{Serber1992,
    author = {Serber, Robert},
    title = {{The Los Alamos Primer: The First Lectures on How to Build an Atomic Bomb}},
    year = {1992},
    publisher = {University of California Press},
    address = {Berkeley, CA, USA}
}

@article{JonesVonHippel1997,
  author = {Jones, Suzanne L. and von Hippel, Frank N.},
  title = "{Transparency measures for subcritical experiments under the CTBT}",
  journal = {Science and Global Security},
  volume = {6},
  number = {3},
  pages = {291-310},
  year = {1997},
  doi = {10.1080/08906139708426410}
}

@article{Anderson_2023,
doi = {10.1088/1748-0221/18/02/P02009},
url = {https://doi.org/10.1088/1748-0221/18/02/P02009},
year = {2023},
month = {feb},
publisher = {IOP Publishing},
volume = {18},
number = {02},
pages = {P02009},
    author = "Anderson, T. and others",
title = {Eos: conceptual design for a demonstrator of hybrid optical detector technology},
journal = {Journal of Instrumentation}
}

@article{Stroujkova:2016,
    author = {Stroujkova, Anastasia and Carnevale, Mario and Vorobiev, Oleg},
    title = {Cavity Radius Scaling for Underground Explosions in Hard Rock},
    journal = {Bulletin of the Seismological Society of America},
    volume = {106},
    number = {6},
    pages = {2500-2510},
    year = {2016},
    month = {09},
    issn = {0037-1106},
    doi = {10.1785/0120160122},
    url = {https://doi.org/10.1785/0120160122},
    eprint = {https://pubs.geoscienceworld.org/ssa/bssa/article-pdf/106/6/2500/2643634/2500.pdf},
}

@article{Rogers:1970, 
 title={Estimating the size of the cavity and surrounding failed region for underground nuclear explosions from scaling rules}, 
 abstractNote={The fundamental physical principles involved in the formation of an underground cavity by a nuclear explosion and breakage of the rock surrounding the cavity are examined from the point of view of making preliminary estimates of their sizes where there is a limited understanding of the rock characteristics. Scaling equations for cavity formation based on adiabatic expansion are reviewed and further developed to include the strength of the material surrounding the shot point as well as the overburden above the shot point. The region of rock breakage or permanent distortion surround ing the explosion generated cavity is estimated using both the Von Mises and Coulomb-Mohr failure criteria. It is found that the ratio of the rock failure radius to the cavity radius for these two criteria becomes independent of yield and dependent only on the failure mechanics of the rock. The analytical solutions developed for the Coulomb-Mohr and Von Mises criteria are presented in graphical form. (author)}, 
 journal={Symposium on Engineering with Nuclear Explosives},
 author={Rogers, Leo A.}, 
 year={1970}, 
 month={May}, 
 pages={p. 519–544} 
 }

@book{Glasstone:1977,
  title={The Effects of Nuclear Weapons},
  author={Glasstone, Samuel and Dolan, Philip J},
  year={1977},
  publisher={U.S. Department of Defense and Energy},
  address={Washington, D.C.},
  edition={3rd}
}

\end{document}